# THE CONCEPTUAL IDEA OF ONLINE SOCIAL MEDIA SITE(SMS) USER ACCOUNT PENETRATION TESTING SYSTEM

Sabarathinam Chockalingam[1], Harjinder Singh Lallie[2]

[1]Kailash Nagar, First Street, Near Police Colony, Karaikudi – 630002, India.
[2]University of Warwick, WMG, Coventry, United Kingdom, CV4 7AL.

**ABSTRACT**

*Social Media Site (SMS) usage has grown rapidly in the last few years. This sudden increase in SMS usage creates an opportunity for data leakage which could compromise personal and/or professional life. In this work, we have reviewed traditional penetration testing process and discussed the failures of traditional penetration testing process to test the 'People' layer of Simple Enterprise Security Architecture (SESA) model. In order to overcome these failures, we have developed the conceptual idea of online SMS user account penetration testing system that could be applied to online SMS user account and the user account could be categorised based on the rating points. This could help us to avoid leaking information that is sensitive and/or damage their reputation. Finally, we have also listed the best practice guidelines of online SMS usage.*

**KEYWORDS**

*Data Leakage, Loss of Reputation, Penetration Testing, Sensitive Information, Social Media Site*

## 1. INTRODUCTION

The lack of awareness of SMS usage best practice in the recent years among the people in an organisation have caused major damages to their professional and/or personal life. This is mainly due to the leakage of sensitive information through their online SMS activity. This introduces the need for the development of *penetration testing* system to test the 'people' layer of the Simple Enterprise Security Architecture (SESA) model, and listing various best practices of online SMS usage.

This work is structured as follows: It begins by reviewing traditional penetration testing process and highlighting the failures of traditional penetration testing process to test the 'people' layer of the SESA model, reviewing various media reports that highlight data leakage through the online SMS activity of Authoritative Persons (APs), and reviewing various problems and challenges relating to online SMS usage particularly in relation to its link with data leakage. It proceeds to devise the conceptual idea of online SMS user account penetration testing system that applies a rating to a number of features of the online SMS user account. It concludes by listing various best practices of online SMS usage.

## 2. LITERATURE REVIEW

This section reviews traditional penetration testing process and highlight the failures of traditional penetration testing methodologies to test the 'People' layer of the SESA model.

### 2.1. Traditional Penetration Testing Process

SESA model encompasses security into three simple layers as shown in Figure 1. They were: Business, People, and System [1]. We have considered SESA model throughout this work because this model is clearly layered with regard to security in an organisation and easy to understand.

Auditing processes typically test the 'Business' layer of the SESA model. 'System' layer comprises of four sub – layers namely, user security interfaces, protocol, code, and component as shown in Figure 1 [1]. Traditional penetration tests test the 'System' layer of the SESA model.

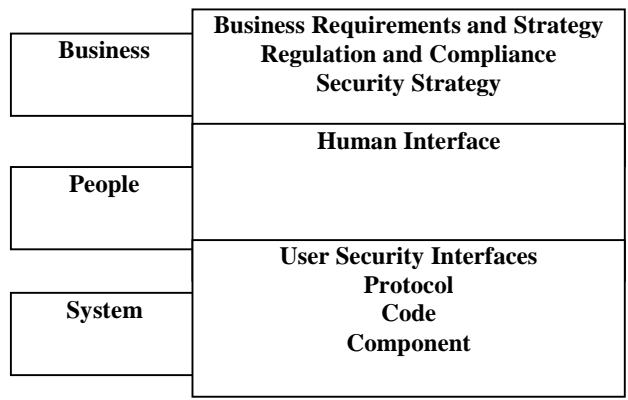

Figure 1. A Simple Enterprise Security Architecture (SESA) Model [1]

Traditional penetration testing process comprises of six phases as shown in Figure 2. They are:

**I. Information Gathering** – This phase would help in gathering information about the system. This information would be used to create a threat model using various techniques [2].

**II. Creating a Threat Model** – This phase would help in creating a threat model which should contain detailed, written description about the major threats to the system ([3] and [4]). Threat scenarios could also be created using sequence diagrams [4].

**III. Building a Test Plan** – This phase would help in building a roadmap for security testing. This should include a high level overview of the test cases, how testing would be conducted, and what would be tested [3].

**IV. Executing Test Cases** – This phase would help in conducting security tests.

**V. The Problem Report** – Critical output of any testing process is the *problem report*. This should mainly include exploit scenarios, severity, and reproduction steps [3]. Reproduction steps in the problem report should help a person who would like to test the system by following the steps listed and would be able to reproduce it.

**VI. Post - mortems** – Bugs are corporate assets that should be examined. Post-mortem evaluation help to refine the testing process to ensure that you could find the bugs sooner in the future security testing [3].

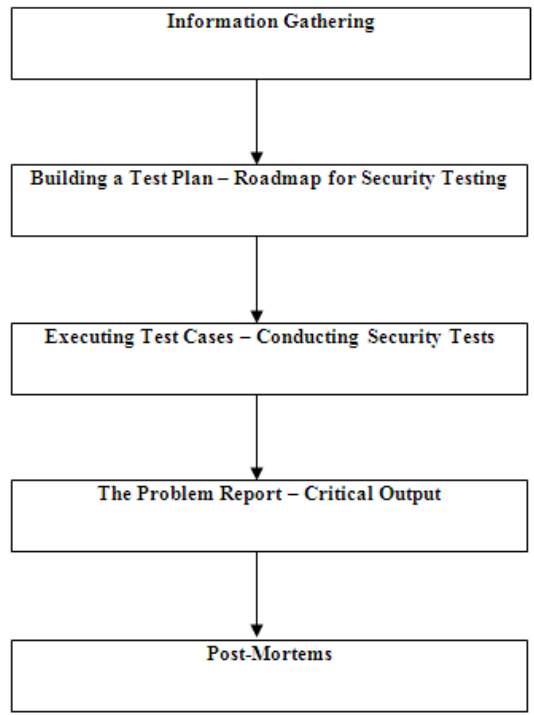

Figure 2. Traditional Penetration Testing Process [3]

'People' layer of the SESA model consist of one sub – layer namely, 'Human Interface' [1].
There are various reasons for the failure of traditional penetration testing process to test the 'people' layer of the SESA model. They are:

**I.** Development of traditional penetration testing process did not consider testing 'people' layer of an organisation. So, It would not be suitable to apply the traditional penetration testing process to test the 'People' layer of the SESA model.

**II.** There are various online SMS used by the people in an organisation. Each online SMS have different features. There is no specific penetration test in the traditional penetration testing process that applies a number of features devised to test online SMS user account. So, It would not be possible to apply the traditional penetration testing process to test the 'people' layer of the SESA model.

## 2.2. Data Leakage Through the Online SMS User Account Activity of the Authoritative Persons

This section reviews various media reports that highlight the data leakage in online SMS user account activity of the Authoritative Persons (APs).

**I.** *"Doctor Booted For Facebook Post – Don't Post That* **[5]***"*

This case states that Dr. Alexandra Thran leaked information about a trauma patient on *Facebook*. This activity on the online SMS compromised her reputation. She also received a reprimand and a $500 fine.

**II.** *"Kevin Pietersen fined by England for obscene Twitter Outburst* **[6]***"*

This case states that Kevin Pietersen posted "F" word on twitter after being left out of the England squad for the ODI and T20 series against Pakistan. Misuse of the online SMS damaged Pietersen's reputation and his relationship with England Cricket Board (ECB). Pietersen have been fined an undisclosed sum for breaching contract.

**III.** *"MI6 Chief Blows his Cover as Wife's Facebook Account Reveals Family Holidays, Showbiz Friends and Links to David Irving* **[7]***"*

This case states that Sir John Sawers (Incoming MI6 head during that time) left exposed by a major personal security breach after his wife posted photos and personal information on *Facebook* such as where they live and work, who their friends are, and where they spend their holidays. This activity by his wife left Sir John Sawers open to criticism and blackmail.

**IV**. *" 'I didn't have the full story' : Ashton Kutcher Forced into Embarrassing Climbdown for Tweeting Joe Paterno Sacking had 'no class'* **[8]***"*

This case states that Ashton Kutcher tweeted "How do you fire Jo Pa? #Insult #Noclass" [8]. Jo Pa, nickname of the legendary football coach who had been fired for allegedly covering up sexual abuse by the school's former football defense coordinator 'Jerry Sandusky'. Ashton Kutcher tweeted this on his official online SMS account after hearing "Joe was fired" without knowing the full story. This online SMS activity damaged Ashton Kutcher's reputation.

These case studies clearly highlight the lack of awareness of the online SMS usage best practice among the Authoritative Persons (APs).

### 2.3 Problems and Challenges of the Online SMS Usage

There are various problems and challenges involved in the online SMS usage particularly in relation to its link with data leakage. They are:

**I. Status/Post Updates** – There are various online SMS in existence. This could be accessed by the user at any time. This allow users to post information / update status as they wish. This highlight that there is a possibility of revealing sensitive information through status updates/ information posted which could compromise their professional and/or personal life [9].

**II. Accepting/Sending Friends' Request** – Carelessness in accepting or sending friends' request could result in adding attackers who would have an easier access to personal information such as date of birth, family details, workplace information, living information, contact information, photos, videos, post/status updates, and friends ([9] and [10]). This could also create an opportunity for the attackers to perform social engineering attacks effectively.

**III. Uploading Photos and Videos** – Everyone could view the publicly available photos and videos of the user. This photos and/or videos might contain sensitive information [9]. This could compromise users' professional and/or personal life.

**IV. Third Party Applications and Links to External Sites** – There is a high likelihood for malware infecting employees' computing platforms when they use the third party applications or click on the external links in the online SMS [9]. This could allow the attackers to monitor and steal intellectual property. There are various privacy issues with using third party applications in the online SMS.

## 3. THE CONCEPTUAL IDEA OF ONLINE SMS USER ACCOUNT PENETRATION TESTING SYSTEM

This section focusses on devising the conceptual idea of online SMS user account penetration testing system that applies a rating to a number of features of the online SMS user account.

### 3.1. Penetration Testing System

This penetration testing system could be applied to Facebook, Twitter, and Linkedin SMS user account. It is impractical to devise a penetration testing system that applies to all SMS because there are various SMS in existence and each SMS have different features. We have chosen Facebook, Twitter, and Linkedin in this work because of its world - wide popularity.
List of features which would be tested in the online SMS user account have been listed below with corresponding description and rating points.

**I. User Name**

**Description:** If the user have same 'username' for Facebook, Twitter, and Linkedin online SMS then it would be easier for the attackers to search and identify user accounts in all these online SMS platforms by knowing it in any one the online SMS platform. This would also give an opportunity for attackers to validate the information they have gathered in one of the online SMS with the other.

**Rating Points:**

**10 Points –** The user have same 'username' for Facebook, Twitter, and Linkedin online SMS.
**5 Points –** The user have same 'username' for two online SMS mentioned above.
**0 Points –** The user have different 'username' for Facebook, Twitter and Linkedin online SMS.

**II. Personal Information**

**Description:** Personal information that could be posted by the user on their online SMS account includes date of birth, family details, workplace information, living information, contact information, and relationship status.

**Rating Points:**

**10 Points** – The user have posted their date of birth, living information, family details, or contact information publicly. These information were given 10 points based on their criticality.
**5 Points** – The user have posted real workplace and/or education information publicly.
**0 Points** – The user have not posted any of their personal information publicly.

### III. Friends List/Connections Visibility

**Description:** Friends/connections of the user are publicly available. This would make it easier for the attackers to gather crucial information about friends/connections.

**Rating Points:**

**10 Points –** Friends/connections of the user are publicly available.
**0 Points** -  Friends/connections of the user are not publicly available.

### IV. Photos/Videos Visibility

**Description:** Photos/videos of the user are publicly available. This would create an opportunity for data leakage. The attackers could view the publicly available photos/videos of the user. This would makes it easier for the attackers to gather sensitive information such as location, car registration number, etc.

**Rating Points:**

**10 Points -** Photos/videos of the user are publicly available.
**0 Points -** Photos/videos of the user are not publicly available.

### V. Photos/Videos

**Description:** Photos/videos posted by the user could leak sensitive information / damage their reputation. Photos/videos posted could contain sensitive information such as date, location, where the photos/videos have been taken, car registration number, friends tagged in the photos/videos, etc.

**Rating Points:**

**10 Points –** Photos/videos posted by the user that contain sensitive information/could damage their reputation.
**0 Points –** There is no photos/videos posted by the user that could leak sensitive information/damage their reputation.

### VI. Posts/Status Updates/Tweets

**Description:** Posts/Status Updates/Tweets posted by the user could leak sensitive information/damage their reputation.

**Rating Points:**

**10 Points –** Posts/Status Updates/Tweets posted by the user that contain sensitive information/damage their reputation.

**0 Points –** There is no posts/Status Updates/Tweets posted in their user account that could leak sensitive information/damage their reputation.

### VII. Photos/Videos Tagged

**Description:** Photos/Videos in which the user is being tagged could leak sensitive information/damage their reputation. Photos/Videos in which the user is being tagged could contain sensitive information such as date, location where photos/videos have been taken, car registration number, other friends being tagged, etc.

**Rating Points:**

**10 Points –** Photos/Videos in which the user is being tagged that could leak sensitive information/damage their reputation.

**0 Points –** There is no such Photos/Videos in which the user is being tagged that could leak sensitive information/damage their reputation.

### VIII. Posts/Status Updates/Tweets Tagged

**Description:** Posts/Status updates/Tweets in which the user is being tagged could leak sensitive information/damage their reputation.

**Rating Points:**

**10 Points –** Posts/Status updates/Tweets in which the user is being tagged could leak sensitive information/damage their reputation.
**0 Points –** There is no such Posts/Status updates/Tweets in which the user is being tagged that could leak sensitive information/damage their reputation.

### IX. Groups/Pages

**Description:** The user have joined any group/page that could damage their reputation.

**Rating Points:**

**10 Points –** The user have joined a group/page that could damage their reputation.
**0 Points** – There is no such group/page that could damage their reputation.

### X. Check – in

**Description:** Facebook feature named 'check-in' could help users to share information about the places they visit at any point of time. This could leak sensitive information about their location.

**Rating Points:**
**10 Points –** The user have used 'check-in' feature in their user account.
**0 Points** – The user have not used 'check-in' feature in their user account.

**XI. Events**

**Description:** In Facebook, users could share publicly about the events they are going attend. This could leak sensitive information such as location, date and time of the event that they are going to attend.

**Rating Points:**

**10 Points –** The user have shared publicly about the events they are going to attend.
**0 Points** – The user have not shared publicly about the events they are going to attend.

This penetration testing system could be applied to the online SMS user account. Based on the points obtained, online SMS user account that have been applied would be categorised as low risk, medium risk, or high risk online SMS user account as shown in Figure 3.

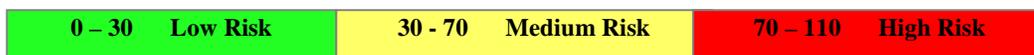

Figure 3. Online SMS User Account Categorisation

**I. Low Risk Online SMS User Account (0 – 30 Points)** – The online SMS User Account contain no or less information that is sensitive/damage user's reputation. The user is recommended to maintain their user account at this category.

**II. Medium Risk Online SMS User Account (30 – 70 Points) –** The online SMS User Account contain reasonable amount of information that is sensitive/damage user's reputation. The user is recommended to take an initiative to review their user account and remove the information that is sensitive/damage their reputation.

**III. High Risk Online SMS User Account (70 – 110 Points) –** The online SMS User Account contain large amount of the information that is sensitive/damage user's reputation. The user is recommended to take necessary actions immediately after reviewing their user account and remove the information that is sensitive/damage their reputation.

The conceptual idea of online SMS user account penetration testing system have been developed taking into account the failure of traditional penetration testing process. This involve three steps as shown in Figure 4.

**I. Information Gathering –** This step would help us to gather information from the online SMS user account which we intend to test.

**II. Applying Gathered Information to the Developed Penetration Testing System –** Once the information have been gathered from the online SMS user account, it would be applied to the developed penetration testing system.

**III. Report Generation –** The final step of the online SMS penetration testing process is the report generation. This report include description, points, and respective category of the online SMS user account based on their points.

International Journal of Security, Privacy and Trust Management (IJSPTM) Vol 3, No 4, August 2014

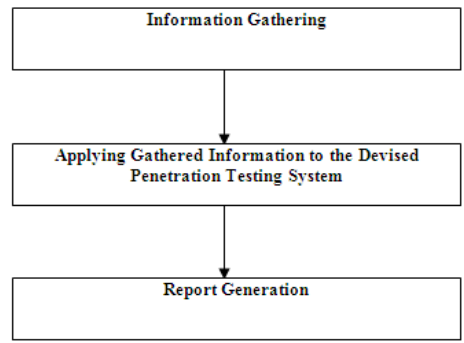

Figure 4. Online SMS Penetration Testing Process

## 4. CONCLUSION

The conceptual idea of online SMS user account penetration testing system have been developed and applied to the online SMS user account of four publicly available Facebook user accounts. Due to privacy issues, we could not share the test process and test report with this paper. Based on the results, we have developed the guidelines to best practice of the online SMS usage.

- ➔ Use different 'username' for the online SMS such as Facebook, Twitter, Linkedin, etc. This would make it difficult for the attackers to search and identify user accounts in various online SMS.
- ➔ Do not post personal information such as date of birth, family details, workplace information, living information, contact information, and relationship status in the online SMS.
- ➔ Do not make friends list/connections visible to everyone in the online SMS.
- ➔ Do not make photos/videos that have been posted by the user visible to everyone in the online SMS.
- ➔ Do not post photos/videos that could leak sensitive information/ damage their reputation.
- ➔ Do not post posts/status updates/tweets that could leak sensitive information/damage their reputation.
- ➔ Do turn on 'Tag Review', which would wait for your approval when your friends' tag you on photos, videos, and posts/status updates.
- ➔ Do not use check – in feature while you are at that particular location. This could leak information about the location you are in at the moment to the attackers easily.
- ➔ Do not join group/page that could damage your reputation.
- ➔ Do not share information publicly about the events that you are going to attend.

## 5. FUTURE WORK

In the future,

- ➔ The conceptual idea of the developed online SMS penetration testing system could be automated.
- ➔ The features of various other online SMS platforms could be considered and incorporated into the developed conceptual idea of the online SMS penetration testing system as it is limited to Facebook, Linkedin, and Twitter.

**Authors**

Sabarathinam Chockalingam (B.Tech., MSc) received his Bachelor of Technology in Computer Science and Engineering from SRM University in 2012, and his Master of Science in Cyber Security and Management from the University of Warwick (Warwick Manufacturing Group) in 2013. After completing his Master of Science, he worked as a Security Researcher – Intern at the Satellite Applications Catapult Limited which involved reviewing various existing security approaches and analysing the feasibility of those approaches with the Internet of Things (IoT). His research interests include Security, Digital Forensics, and Data Analytics.

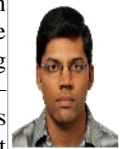

Harjinder Singh Lallie(BSc., MSc., MPhil) is a senior teaching fellow in Cybersecurity at the University of Warwick (WMG). He teaches three modules on the programme. He has developed and led a number of very successful University courses in Digital Forensics and Security at both undergraduate and postgraduate level.His research focus is in the area of Digital Forensics and Information Security particularly focussing on social network analysis and is currently studying towards his PhD. He has published dozens of papers in the digital forensics/information security domain and has presented at numerous conferences/workshops. He has held a number of conference committee memberships and acts as an external examiner for three Universities. Harjinder is a respected academic in the area of Teaching, Learning, Assessment and Curriculum (TLAC) and regularly organises and delivers at workshops and conferences in this domain.

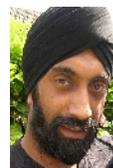